\documentclass[aps, prd, twocolumn, superscriptaddress]{revtex4-1}
\usepackage{graphicx}
\usepackage{amssymb}
\usepackage{amsmath}
\usepackage{bm}
\usepackage{subfigure}
\usepackage[colorlinks=True,linkcolor=blue,anchorcolor=blue,citecolor=blue,CJKbookmarks=true]{hyperref}

\begin{document}
\title{Enhancing pair production with optimized chirped laser fields}

\author{Z. L. Li }
\email{zlli@cumtb.edu.cn}
\affiliation{School of Science, China University of Mining and Technology, Beijing 100083, China}

\author{Y. F. Chen}
\affiliation{School of Science, China University of Mining and Technology, Beijing 100083, China}

\author{R. Z. Jiang}
\affiliation{State Key Laboratory for Tunnel Engineering, China University of Mining and Technology, Beijing 100083, China}

\author{Y. J. Li}
\email{lyj@aphy.iphy.ac.cn}
\affiliation{School of Science, China University of Mining and Technology, Beijing 100083, China}
\affiliation{State Key Laboratory for Tunnel Engineering, China University of Mining and Technology, Beijing 100083, China}

\date{\today}

\begin{abstract}
The optimal chirped field for enhancing electron-positron (EP) pair production is explored using a quantum kinetic approach. First, the momentum spectrum and number density of EP pairs produced by Gaussian chirped fields are investigated. The results show that the momentum spectrum exhibits distinct interference patterns, while the number density grows monotonically with chirp parameters but oscillates with the carrier angular frequency. Moreover, the number density increases by four orders of magnitude compared to chirp-free fields. The results are further compared with those from four other chirped fields: frequency-modulated, linear, quadratic, and sinusoidal chirp. The analysis reveals that the maximum number density for sinusoidally chirped fields is the highest, followed by Gaussian, frequency-modulated, quadratically, and linearly chirped fields. This ranking also applies to the maximum enhancement factors for these chirped fields. Notably, the number density for sinusoidally chirped fields improves nine orders of magnitude compared to chirp-free fields. These results not only deepen our understanding of pair production in chirped fields but also provide significant optimization strategies for future vacuum pair production experiments.
\end{abstract}

\maketitle

\section{INTRODUCTION}

Since Dirac proposed the hole theory to solve the difficulty of negative energy solutions for the relativistic wave equation, we have gained a deeper understanding of vacuum structure \cite{Dirac1928}.
The vacuum is not devoid of matter.
It contains a sea of virtual particle pairs undergoing constant formation and destruction.
The virtual particle pairs can be separated by ultrastrong external fields, leading to the production of real particle-antiparticle pairs \cite{Sauter1931,Heisenberg1936,Schwinger1951}.
Electron-positron (EP) pairs produced from vacuum by a strong background field is a famous theoretical prediction of quantum electrodynamics (QED) \cite{Dunne2009,Piazza2012,Xie2017,Fedotov2023}.
To create detectable EP pairs by an electric field, the field strength must reach the critical value $E_{\mathrm{cr}}=m^2/e\approx 1.3\times10^{18}\mathrm{V/m}$, where $m$ is the electron mass and $e$ represents the magnitude of the electron charge.
Note that the natural unit $\hbar=c=1$ is used in this paper.
The laser intensity corresponding to the critical field strength is  $I_{\mathrm{cr}}\approx 4.3\times 10^{29}\mathrm{W/cm^2}$.
With the rapid development of laser technology, especially the invention of chirped pulse amplification (CPA) technology \cite{Strickland1985}, laser intensity has been greatly improved. At present, the laser intensity has achieved around $10^{23}\mathrm{W/cm^2}$.
However, this is still far below the critical laser intensity.
Therefore, vacuum pair production has not yet been experimentally validated.

In order to observe this process in experiments, researchers have been striving to continuously increase the laser intensity \cite{ELI,XCEL}, while also exploring methods to lower the pair production threshold or improve the particle yield \cite{Schutzhold2008,Bell2008,DunnePRD2009,Piazza A2009,Bulanov2010,Titov2012,Abdukerim2012,Li2014,Torgrimsson2016, Schneider2016,Torgrimsson2019,Kohlf2021,Li-2021}.
One of the effective methods to increase the particle yield is the dynamically assisted Schwinger mechanism \cite{Schutzhold2008}.
The superposition of a high-frequency weak field onto a low-frequency strong field significantly enhances the production of EP pairs.
This mechanism can be interpreted in two ways: the weak field effectively shortens the tunneling distance required for pair production in the strong-field regime, or the strong field opens additional pair production channels that are normally inaccessible under weak-field conditions.
Another approach to enhance pair production is to use chirped fields \cite{Dumlu2010,Min2013,Abdukerim2017,Olugh2019,Ababekri2020,
Gong2020,Wang2021,Mohamedsedik2021,Li2021,Xie2022,Osman2023,Chen2024}.
By choosing suitable chirp parameters, the number density can be improved several orders of magnitude.
This is primarily because the chirp increases the high-frequency components of the field, enhancing pair production through multiphoton absorption.
Furthermore, the dynamically assistant mechanism combining high-frequency and low-frequency components also contributes significantly.
Currently, extensive research has been conducted on pair production under chirped electric fields with different chirp profiles (linear, quadratic, frequency-modulated, and sinusoidal), various polarization states (linear, elliptical, and circular), and spatial homogeneity or inhomogeneity.
However, existing studies have mainly focused on examining maximum particle yields and enhancement factors for individual chirped fields.
The comparative optimization, determining which type of chirped field can maximize particle yield and enhancement factor under achievable field parameters in the future, remains unexplored.
In this paper, we will study this unexplored territory using quantum Vlasov equation (QVE). We first study the momentum spectrum and number density of EP pairs created by a Gaussian chirped electric fields. Then within a reasonable range of field parameters, we compare the number density for the Gaussian chirped field with that for the other four different chirped fields to identify the optimal chirped field that maximizes particle yield. In addition, the enhancement factors for these five chirped fields are also calculated. Their maximum values are compared to identify the optimal chirped field that maximizes the enhancement factor. This study will serve as a valuable reference for advancing the optimal control theory of EP pair production \cite{Kohlfust2013,Hebenstreit2014,Dong2020} and for experimentally validating the production of EP pairs.

This paper is organized as follows: In Sec.~\ref{sec:two} we briefly presented the QVE approach, which facilitates the computation of the momentum spectrum and number density. In Sec.~\ref{sec:three}, the momentum spectrum and number density of EP pairs created by a Gaussian chirped electric field with different chirp parameters are investigated.
The number density and enhancement factors for five different chirped electric fields are also calculated numerically. Their maximum values are compared respectively to identify the optimal chirp field that produces the maximum number density and enhancement factor. Section~\ref{sec:four} is a summary and outlook.

\section{Theoretical model and method}
\label{sec:two}

In this study, we consider EP pair production by a time-dependent and spatially homogeneous electric field, $\mathbf{E}(t)=(0,0,E(t))$, which can be regarded as the field at the antinode of the stationary wave formed by two counter-propagating laser fields.
The corresponding potential in temporal gauge is $\mathbf{A}(t)=(0,0,A(t))$, where $E(t)=-\dot{A}(t)$.

Starting from Dirac equation
\begin{eqnarray}\label{dirac}
 [i\gamma^{0}\frac{\partial}{\partial t}+i\bm{\gamma}\cdot[\nabla_\mathbf{x}-ie\mathbf{A}(t)]-m]\Psi(\mathbf{x},t)=0
\end{eqnarray}
and employing a canonical time-dependent Bogoliubov transformation, we can derive the quantum Vlasov equation:
\begin{eqnarray}\label{eq:qve}
\dot{f}(\mathbf{p},t)\!=\!\frac{1}{2}\lambda(\mathbf{p},t)\!\!
 \int^{t}_{\!-\!\infty}\!\!dt'\lambda(\mathbf{p},t')[1\!-\!2f(\mathbf{p},t')]
 \cos\theta (t,t'),\quad
\end{eqnarray}
where $f(\mathbf{p},t)$ is the momentum distribution function, $\lambda(\mathbf{p},t)=eE(t)\varepsilon_{\perp}/\varepsilon^{2}(\mathbf{p},t)$ , and $\theta(t,t')=2\int^{t}_{t'}dt'' \varepsilon(\mathbf{p},t'')$ is the dynamic phase. $\mathbf{p}=(\mathbf{p}_{\bot},p_{\parallel})$ denotes the canonical momentum, $\varepsilon(\mathbf{p},t)=\sqrt{\varepsilon_{\bot}^{2}
+k_{\parallel}^{2}(t)}$ represents the energy of particles,  $\varepsilon_{\bot}=\sqrt{m^{2}+\mathbf{p}_{\bot}^{2}}$ is the perpendicular energy, and $k_{\parallel}(t)=p_{\parallel}-eA(t)$ is the kinetic momentum.

In order to solve Eq. (\ref{eq:qve}) numerically, two auxiliary variables are introduced by
 \begin{eqnarray}\label{u}
  u(\mathbf{p},t)=\int^{t}_{-\infty}dt'\lambda(\mathbf{p},t')[1-2f(\mathbf{p},t')]\cos\theta (t,t'),
 \end{eqnarray}
 \begin{eqnarray}\label{v}
  v(\mathbf{p},t)=\int^{t}_{-\infty}dt'\lambda(\mathbf{p},t')[1-2f(\mathbf{p},t')]\sin\theta (t,t'),
 \end{eqnarray}
so Eq. (\ref{eq:qve}) can be equivalently transformed into a first-order ordinary differential equation system:
\begin{eqnarray}\label{eq:system}
\dot{f}(\mathbf{p},t)&=&\frac{1}{2}\lambda(\mathbf{p},t)u(\mathbf{p},t),
 	\nonumber \\ {\dot{u}(\mathbf{p},t)}&=&\lambda(\mathbf{p},t)[1-2f(\mathbf{p},t)]
 -2\varepsilon(\mathbf{p},t)v(\mathbf{p},t),\; \\ {\dot{v}(\mathbf{p},t)}&=&2\varepsilon(\mathbf{p},t)u(\mathbf{p},t)
 .\nonumber
 \end{eqnarray}
Since the system initially is in a field-free vacuum state, the initial condition is $f(\mathbf{p},-\infty)$ = $u(\mathbf{p},-\infty)$ = $v(\mathbf{p},-\infty)=0$. The distribution function $f(\mathbf{p},t)$ can be obtained by solving Eqs. (\ref{eq:system}). Integrating the distribution function $f(\mathbf{p},+\infty)$ over momentum results in the final number density of created pairs:
 \begin{equation}
 n(+\infty)=2\int\frac{d^3\mathbf{p}}{(2\pi)^3}f(\mathbf{p},+\infty),
 \end{equation}
where $2$ comes from the spin degeneracy.

\section{Numerical Results}
\label{sec:three}

This section is divided into two parts. The first part examines the momentum distribution and number density of created EP pairs within a Gaussian chirped electric field. The second one calculates the number density of EP pairs created by five different chirped electric fields as a function of chirp parameters and carrier angular frequency, aiming to find the optimal chirped field for maximizing the particle yield and the gain in pair production.

\subsection{Gaussian chirped electric fields}
\label{sub:1}

The Gaussian chirped electric field considered in this part has the form
\begin{equation}\label{eq:filed1}
 E_G(t)=E_{0} \exp\Big(-\frac{t^{2}}{2\tau^{2}}\Big) \cos\Big(\omega_{0} t+b\exp\Big(-\frac{t^2}{2\omega_{m}^2}\Big)t\Big),
\end{equation}
where $E_{0}$ is the field strength, $\omega_{0}$ is the carrier angular frequency, and $\tau$ denotes the pulse duration. $b$ and $\omega_{m}$ are the chirp parameters. The Gaussian chirped electric field Eq. (\ref{eq:filed1}) can be rewritten as
\begin{equation}\label{eq:filedgx}
E_G(t)=E_{0}\exp\Big(-\frac{t^{2}}{2\tau^{2}}\Big)\cos(\omega_\mathrm{eff}t),
\end{equation}
where $\omega_\mathrm{eff}=\omega_{0}+b\exp(-\frac{t^2}{2\omega_{m}^2})$ is a time-dependent effective frequency associated with the chirp parameters $b$ and $\omega_{m}$. To ensure that the frequency modulation is within a reasonable range of field parameters, we constrain that the effective frequency at any given time cannot exceed twice the carrier angular frequency, namely $|b\exp(-\frac{t^2}{2\omega_{m}^2})|\leq\omega_{0}$ for any time $t$. Since $|b\exp(-\frac{t^2}{2\omega_{m}^2})|_{\mathrm{max}}=b$, we get the relationship $b\leq\omega_{0}$.

It is well-known that there are two main mechanisms for vacuum pair production in pure laser fields. One is multiphoton absorption, and the other is the Schwinger effect (tunneling mechanism). These two mechanisms can distinguished by the Keldysh adiabaticity parameter  $\gamma=m\omega/|e|E_{0}$~\cite{Keldysh1965,Brezin1970}. For the former, $\gamma\gg1$, and for the latter, $\gamma\ll1$. However, in this paper we focus on the more interesting scope
$\gamma\sim\mathcal{O}(1)$. Considering the laser field that can be achieved in the future, the field parameters are chosen as $E_{0}=0.1E_{\mathrm{cr}}$, $\omega_{0}=0.1\sim0.6m$, and $\tau=20/m$.

\begin{figure}[!ht]
\centering
\includegraphics[width=0.48\textwidth]{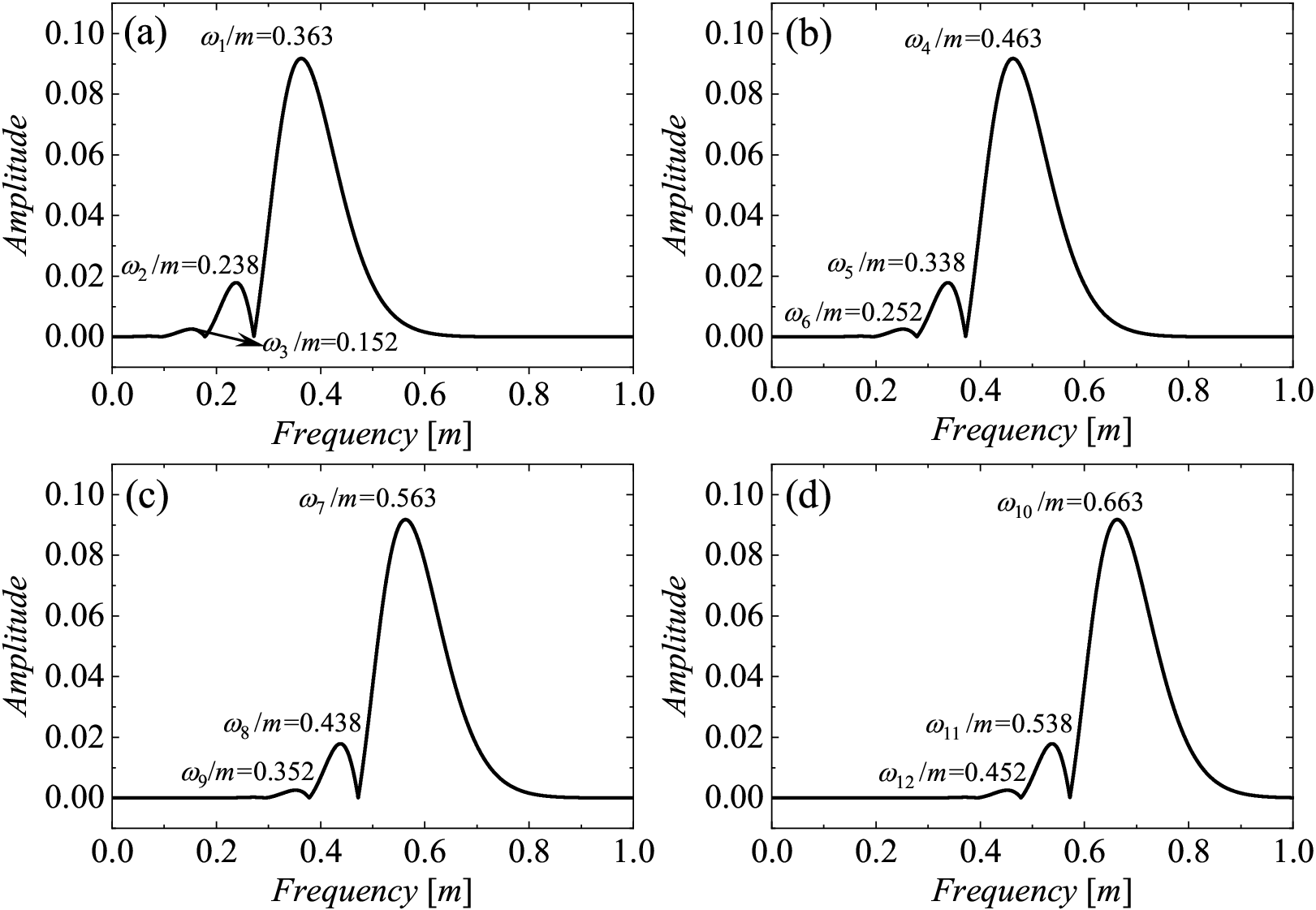}
\caption{The frequency spectra of the Gaussian chirped electric fields for different values of the carrier angular frequency $\omega_0$. (a) $\omega_{0}=0.3m$ , (b)  $\omega_{0}=0.4m$, (c) $\omega_{0}=0.5m$, and (d) $\omega_{0}=0.6m$. The peak positions are also shown in the figures. Other electric field parameters are $E_{0}=0.1E_{\mathrm{cr}}$, $\tau=20/m$, $b=0.1m$, and $\omega_{m}=30/m$. \label{fig:frequency}}
\end{figure}

To quantitatively analyze the peaks in the momentum spectrum of EP pairs created by the Gaussian chirped electric fields, we plot the frequency spectra of the field (\ref{eq:filed1}) for different carrier angular frequencies in Fig. \ref{fig:frequency}. The chirp parameters are $b=0.1m$ and $\omega_{m}=30/m$. In the frequency spectrum, the frequencies corresponding to the peaks are marked. From Fig. \ref{fig:frequency}, one can see that there are three peaks in each frequency spectrum and the carrier angular frequencies are no longer the main frequencies. Because of the frequency modulation caused by the positive chirp parameter $b$, all main peak frequencies are larger than the carrier angular frequency $\omega_{0}$. Moreover, with the increase of the carrier angular frequency, the frequency spectrum remains almost unchanged except for the overall shift towards higher frequencies.

\begin{figure}[!ht]
\centering
\includegraphics[width=0.48\textwidth]{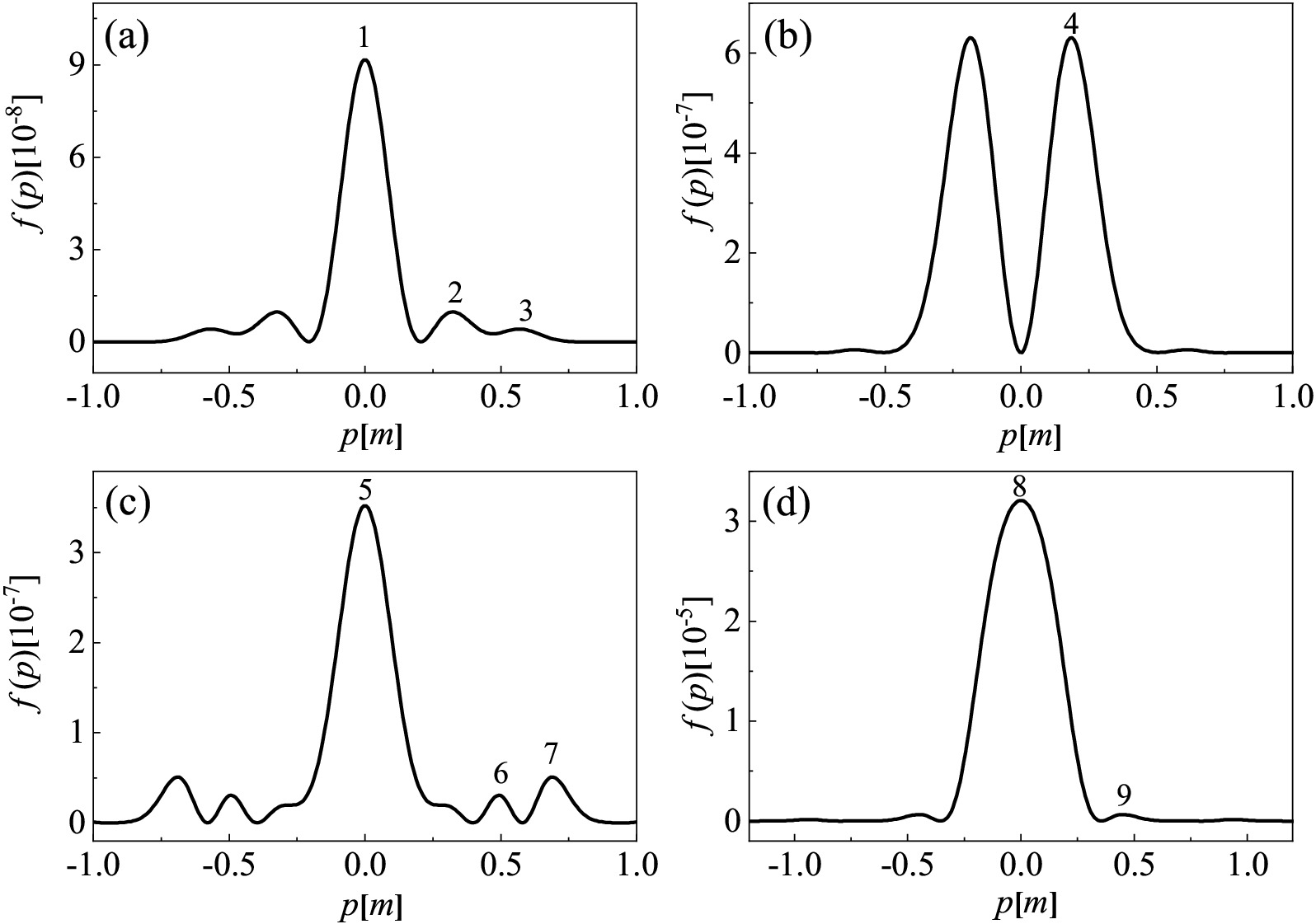}
\caption{The longitudinal momentum spectra ($\mathbf{p}_\perp=0$) of EP pairs produced by Gaussian chirped electric fields for different values of the carrier angular frequency $\omega_0$. (a) $\omega_{0}=0.3m$, (b) $\omega_{0}=0.4m$, (c) $\omega_{0}=0.5m$, and (d) $\omega_{0}=0.6m$. The peaks in the momentum spectra are marked with $1\sim9$. Other electric field parameters are $E_{0}=0.1E_{\mathrm{cr}}$, $\tau=20/m$, $b=0.1m$, and $\omega_{m}=30/m$. \label{fig:ppu}}
\end{figure}

In Fig. \ref{fig:ppu}, the longitudinal momentum spectra ($\mathbf{p}_\perp=0$) of EP pairs created in Gaussian chirped electric fields with different carrier angular frequencies are depicted. The obvious peaks in the momentum spectra are marked with numbers $1\sim9$. As can be seen from the figure, the momentum spectra exhibit distinct interference patterns and the interference patterns vary with carrier angular frequency. To achieve a more precise quantitative understanding of the interference patterns, we utilize the energy conservation equation in multiphoton pair production process,
\begin{equation}\label{eq:enery}
\mathcal{E}(\mathbf{p})=2\sqrt{m_{\ast}^{2}+\mathbf{p}^{2}}=N\omega,
\end{equation}
where $\mathcal{E}(\mathbf{p})$ denotes the total energy of a pair of EP pairs, $N$ represents the number of absorbed photons, $\omega$ is the frequency of absorbed photons, and $m_{\ast}$ is the effective mass of particles as~\cite{Kohlfurst2014}
\begin{equation}\label{eq:m}
m_{\ast}=m\sqrt{1+\frac{e^{2}}{m^{2}}\frac{E_{0}^{2}}{2\omega^{2}}}.
\end{equation}
According to the energy-momentum relation of relativistic particles in Eq. (\ref{eq:enery}), the total energy $\mathcal{E}_{1}$, $\mathcal{E}_{2}$, ... , and $\mathcal{E}_{9}$ corresponding to the momentum at peaks $p_{1}$, $p_{2}$, ... , and $p_{9}$ can be calculated. For example, in Fig. \ref{fig:ppu}(a), the total energy corresponding to $p_{1}=0.001m$, $p_{2}=0.323m$, and $p_{3}=0.567m$ is $\mathcal{E}_{1}=2.038m\approx5\omega_{1}+\omega_{2}$, $\mathcal{E}_{2}=2.138m\approx6\omega_{1}$, and $\mathcal{E}_{3}=2.332m\approx6\omega_{1}+\omega_{3}$, respectively. Note that the effective mass is calculated using the frequency $\omega_1$ in Fig. \ref{fig:frequency}(a), because it is the primary type of photon absorbed. Based on the above results, it is easy to find that the peaks corresponding to $p_{1}$, $p_{2}$, and $p_{3}$ are associated with six-, six-, and seven-photon absorption, respectively. The introduction of chirp enables the six-photon absorption that cannot occur for the electric field without chirp. Thus, it can enhance the pair production significantly. Moreover, for $p_{4}=0.185m$ and $p_{7}=0.639m$, the total energy is $\mathcal{E}_{4}=2.058m\approx3\omega_{4}+2\omega_{5}$ and  $\mathcal{E}_{7}=2.387m\approx3\omega_{7}+2\omega_{9}$. Both are associated with five-photon absorption. For $p_{5}=0$, $p_{6}=0.492m$, $p_{9}=0.452m$, and $p_{8}=0$ in Figs. \ref{fig:ppu}(c) and (d), we can obtain $\mathcal{E}_{5}=2.016m\approx3\omega_{7}+\omega_{9}$, $\mathcal{E}_{6}=2.243m\approx4\omega_{7}$, $\mathcal{E}_{9}=2.205m\approx2\omega_{10}+2\omega_{12}$, and $\mathcal{E}_{8}=2.011m\approx3\omega_{10}$. The first three belong to four-photon absorption, while the last one pertains to three-photon absorption. Consequently, the interference fringes in the momentum spectra actually correspond to multiphoton absorption rings.

\begin{figure}[!ht]
\centering
\includegraphics[width=0.48\textwidth]{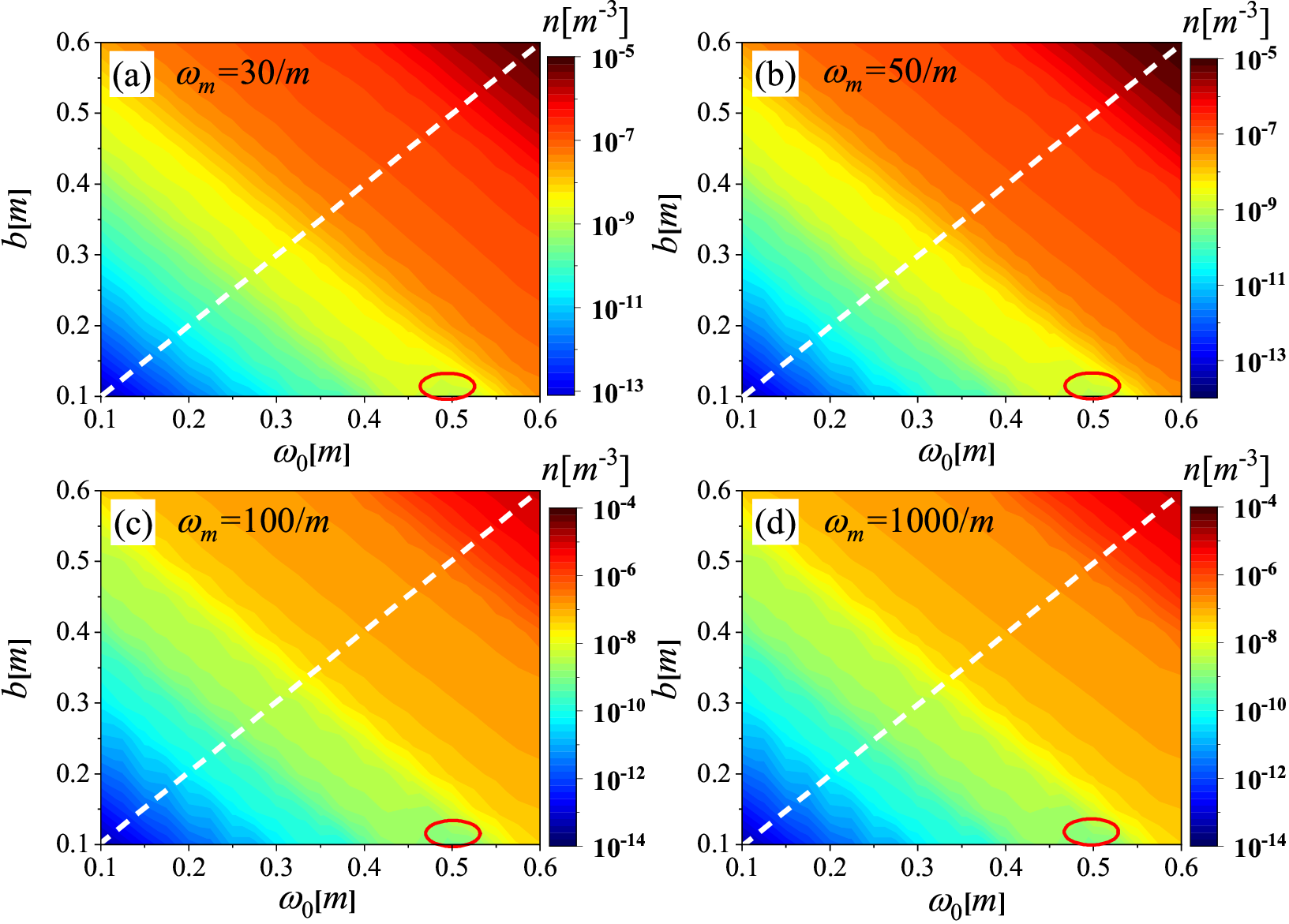}
\caption{The number density of EP pairs created by a Gaussian chirped electric field, as a function of the chirp parameter $b$ and the carrier angular frequency $\omega_{0}$, for different values of the chirp parameter $\omega_m$. (a) $\omega_{m}=30/m$, (b) $\omega_{m}=50/m$, (c) $\omega_{m}=100/m$, and (d) $\omega_{m}=1000/m$. Other electric field parameters are $E_{0}=0.1E_{\mathrm{cr}}$ and $\tau=20/m$.  \label{fig:ent}}
\end{figure}

\begin{figure}[!ht]
\centering
\includegraphics[width=0.48\textwidth]{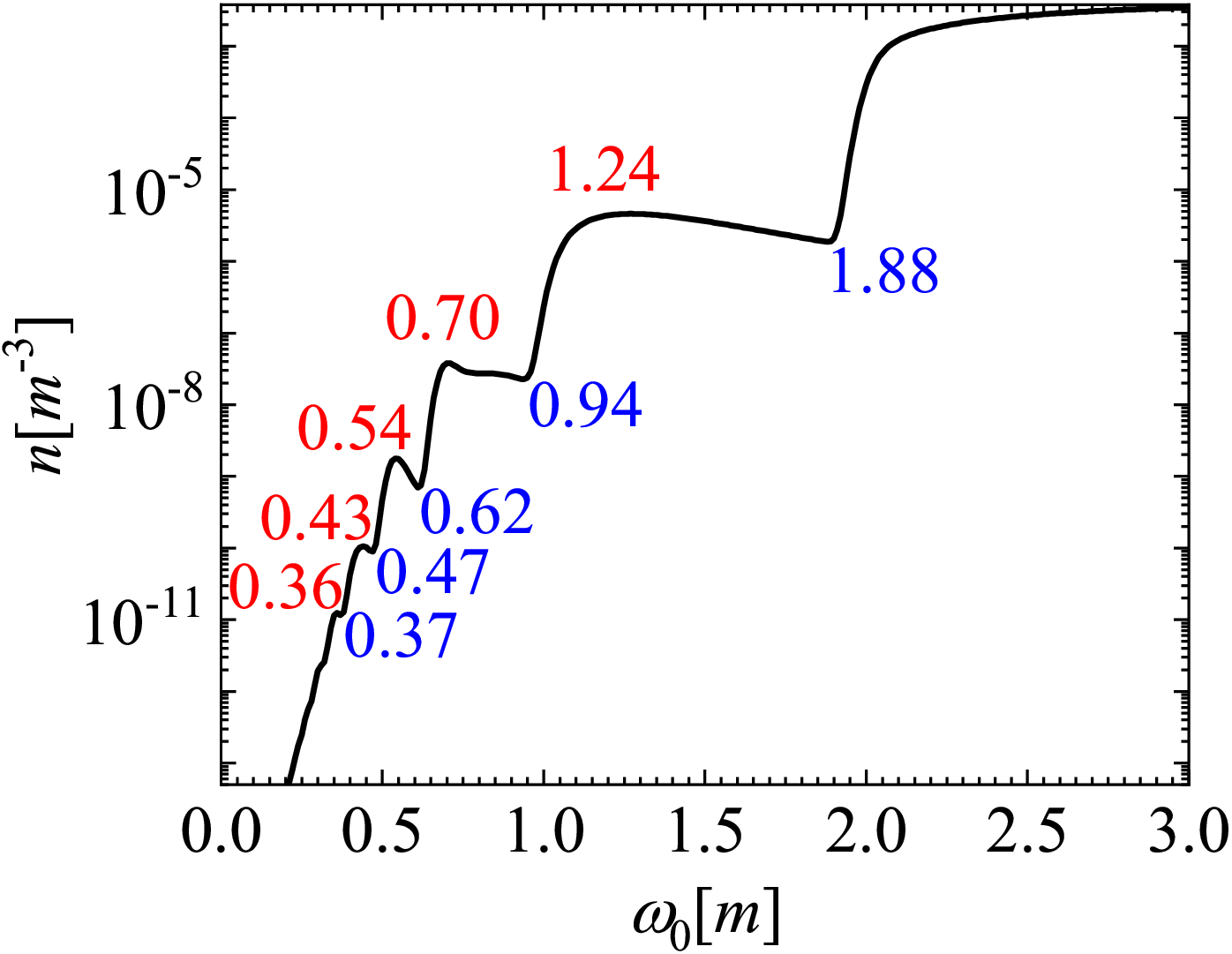}
\caption{The number density of EP pairs produced by a chirp-free electric field ($b=0$) as a function of the carrier angular frequency $\omega_{0}$. The peaks marked in the figure correspond to multiphoton absorption thresholds. Other electric field parameters are $E_{0}=0.1E_{\mathrm{cr}}$ and $\tau=20/m$. \label{fig:b0t}}
\end{figure}

Now we discuss the effect of chirp parameters on the number density of created EP pairs.
According to the constraint of chirp parameters discussed earlier, the maximum value of the chirp parameter $b$ should be less than or equal $\omega_0$ while the chirp parameter $\omega_{m}$ can choose any value except $0$.
In Fig. \ref{fig:ent}, we plot the number density of created EP pairs as a function of the chirp parameter $b$ and the carrier angular frequency $\omega_0$ for $\omega_m=30/m$, $50/m$, $100/m$, and $1000/m$.
Note that due to the constraint of chirp parameters, we only focus on the date on the right half side of the white dashed line in Fig. \ref{fig:ent}.
From the figure, we find that the influence of the chirp parameter $\omega_m$ on the number density is relatively weak.
However, for a given $\omega_{m}$ and $\omega_{0}$, the increase of the chirp parameter $b$ can considerably increase the number density of created EP pairs.
For instance, the maximum increase can reach three orders of magnitude.
Similar to the case of $b$, the number density generally increases with the increase of carrier angular frequency $\omega_{0}$.
The maximum increase can reach six orders of magnitude, whose growth is comparatively more robust than that of $b$.
Notably, from Figs. \ref{fig:ent}(a)-(d), we discover that the data within the red cycle does not follow the growth trend discussed above.
In this region, the number density decreases with the increase of $\omega_0$.
This phenomenon can be elucidated by the relation between the number density and the carrier angular frequency $\omega_{0}$ shown in Fig. \ref{fig:b0t}.
Firstly, we review the relationship between the effective frequency $\omega_\mathrm{eff}$ and the chirp parameters $b$ and $\omega_{m}$: $\omega_\mathrm{eff}=\omega_{0}+b\exp(-\frac{t^2}{2\omega_{m}^2})$. It is evident that the effective frequency is affected by the chirp parameters $b$ and $\omega_{m}$.
When $\omega_{m}$ reaches a sufficiently large value, the effective frequency $\omega_\mathrm{eff}$ approaches $\omega_{0}+b$.
For the data in the red cycle, for example, $\omega_0=0.5m$, the effective frequency $\omega_\mathrm{eff}=0.5m+0.1m=0.6m$, which is about at the minimum value of the number density, see Fig. \ref{fig:b0t}.
Consequently, the number density in this region is lower than that of its adjacent region.
However, from a broader perspective, the number density of created EP pairs increases with the increase of the chirp parameter $b$, $\omega_m$, and the carrier angular frequency.
In the presence of Gaussian chirped electric fields, the maximum value of the number density is $n^{\mathrm{max}}_\mathrm{G}=1.19\times10^{-5}m^{-3}$. The associated electric field parameters are $\omega_{0}=0.6m$, $b=0.6m$, and $\omega_{m}=1000/m$.
This result is four orders of magnitude higher than the number density in chirp-free electric fields.
Therefore, pair production can be significantly enhanced by a Gaussian chirped electric fields.

\subsection{Optimizing chirped electric fields }
\label{sub:2}

In this section, we examine different types of chirped fields to determine which one can maximize the particle yield within a reasonable range of electric field parameters. We also search for the maximum ratio of the number density in chirped and chirp-free electric fields. The optimal electric field parameters are also considered.

The background laser field used is similar to that outlined in the preceding text, characterized as a spatially homogeneous and time dependent electric field. Alongside the Gaussian chirped electric field, we also introduce various other types of chirped fields. They are frequency-modulated chirped electric fields, linearly chirped electric fields, quadratically chirped electric fields, and sinusoidally chirped electric fields. The specific expressions are
\begin{equation}\label{eq:filed2}
E_\mathrm{F}(t)=E_{0} \exp\Big(-\frac{t^{2}}{2\tau^{2}}\Big) \cos(\omega_{0} t+b_{1}\sin(\omega_{n}t)),
\end{equation}
\begin{equation}\label{eq:filed3}
E_\mathrm{L}(t)=E_{0} \exp\Big(-\frac{t^{2}}{2\tau^{2}}\Big) \cos(\omega_{0} t+b_{2}t^2),
\end{equation}
\begin{equation}\label{eq:filed4}
E_\mathrm{Q}(t)=E_{0} \exp\Big(-\frac{t^{2}}{2\tau^{2}}\Big) \cos(\omega_{0} t+b_{3}t^2+b_{4}t^3),
\end{equation}
\begin{equation}\label{eq:filed5}
E_\mathrm{S}(t)=E_{0} \exp\Big(-\frac{t^{2}}{2\tau^{2}}\Big) \cos(\omega_{0} t+b_{5}\omega_{0} t\sin(b_{6}\omega_{0} t)),
\end{equation}
where $b_{1}, b_{2}, b_{3}, b_{4}, b_{5}, b_{6}$, and $\omega_{n}$ denote the chirp parameters. In numerical calculations, these chirped electric fields have the same parameters, i.e.,  $E_{0}=0.1E_{\mathrm{cr}}$, $\omega_{0}=0.1\sim0.6m$, and $\tau=20/m$.

In order to maintain a reasonable effective frequency, we need to impose certain constraints on the chirp parameters.
Note that the opening and closing times of the electric field are $-200/m$ and $200/m$, respectively.
For the frequency-modulated chirped electric field (\ref{eq:filed2}), the effective frequency $\omega_\mathrm{eff}=\omega_{0}+\frac{b_{1}\sin(\omega_{n}t)}{t}$. To ensure the effective frequency at any given time cannot exceed twice the carrier angular frequency, we have $|\frac{b_{1}\sin(\omega_{n}t)}{t}|\leq\omega_{0}$. Because of $|\frac{b_{1}\sin(\omega_{n}t)}{t}|_{\max}=b_{1}\omega_{n}$, we obtain the inequality $b_{1}\omega_{n}\leq\omega_{0}$.
Considering that the impact of $b_1$ is greater than that of $\omega_n$, we constrain the upper limit of $b_{1}$ to 10.
For the linearly chirped electric field (\ref{eq:filed3}), $\omega_\mathrm{eff}=\omega_{0}+b_{2}t$, requiring $|b_{2}t|\leq\omega_{0}$.
Since the existence time of the electric field $|t|\leq200/m$, it is sufficient for requiring $b_{2}\leq m\omega_{0}/200$.
For the quadratically chirped electric field (\ref{eq:filed4}), $\omega_\mathrm{eff}=\omega_{0}+b_{3}t+b_{4}t^2$, requiring $b_{3}t+b_{4}t^2\leq\omega_{0}$.
Let $b_{3}=\beta_{1}\omega_{0}/(2\tau)$ and $b_{4}=\beta_{2}\omega_{0}/(2\tau^2)$, where ($\beta_{1,2}\geq0$), we deduce that $\frac{\beta_{1}}{2}\frac{t}{\tau}+\frac{\beta_{2}}{2}\frac{t^2}{\tau^2}\leq1$. Since $|\frac{t}{\tau}|\leq10$, we can obtain $5\beta_{1}+50\beta_{2}\leq1$.
For the sinusoidally chirped electric field (\ref{eq:filed5}), $\omega_\mathrm{eff}=\omega_{0}+b_{5}\omega_{0} \sin(b_{6}\omega_{0} t)$, requiring $|b_{5}\omega_{0} \sin(b_{6}\omega_{0} t)|\leq\omega_{0}$ for any time $t$.
Due to $|b_{5}\omega_{0} \sin(b_{6}\omega_{0} t)|_{\max}=b_{5}\omega_{0}$, we obtain $b_{5}\leq1$. Furthermore, it is reasonable to require a frequency not exceeding $m$. Then we have $b_{6}\omega_{0}\leq m$, i,e., $b_{6}\leq m/\omega_{0}$.

\subsubsection{Maximizing the particle yield}
Now we will calculate the number density of EP pairs created by different chirped electric fields and find the optimal chirped electric field to maximize the number density.

\begin{figure}[!ht]
\centering
\includegraphics[width=0.48\textwidth]{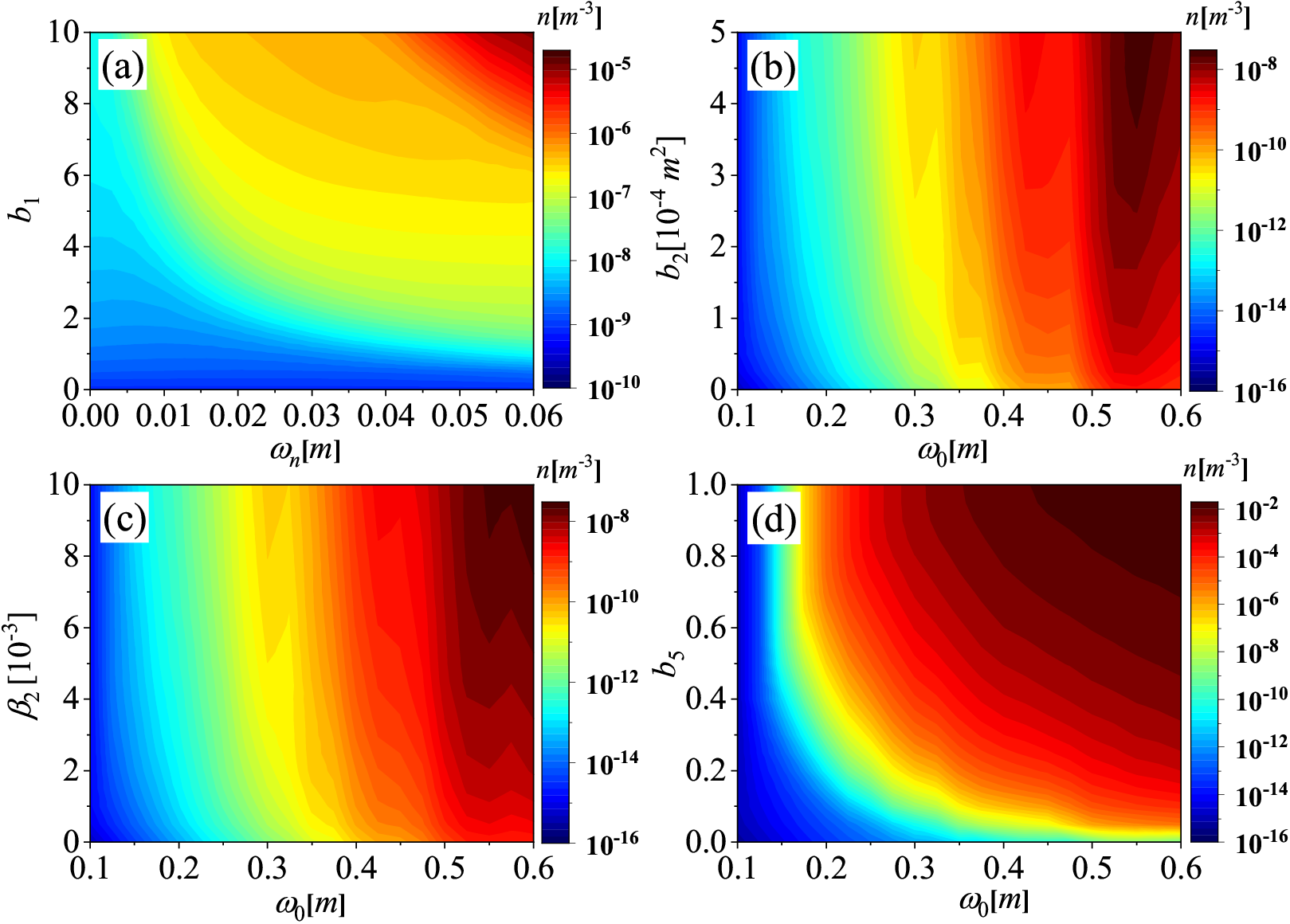}
\caption{The number density of created EP pairs as a function of chirp parameters and the carrier angular frequency for different chirped electric fields. (a) frequency-modulated chirped electric fields with $\omega_{0}=0.6m$, (b) linearly chirped electric fields, (c) quadratically chirped electric field with $\beta_{1}=0.1$, and (d) sinusoidally chirped electric fields with $b_{6}=1.6$. Other electric field parameters are $E_{0}=0.1E_{\mathrm{cr}}$ and $\tau=20/m$. \label{fig:ntmax}}
\end{figure}

\begin{figure}[!ht]
\centering
\includegraphics[width=0.48\textwidth]{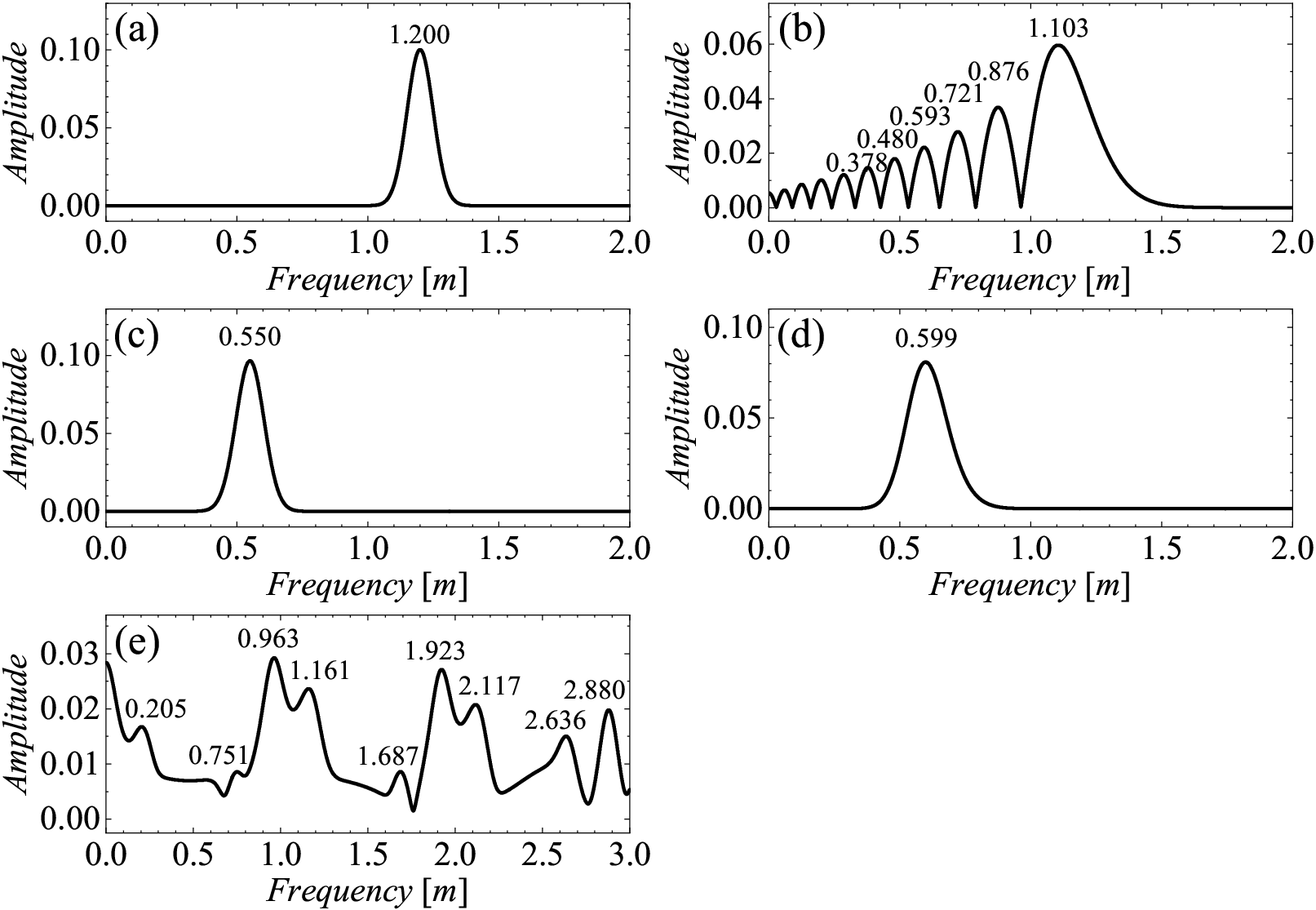}
\caption{The frequency spectrum of five different chirped electric fields at optimal field parameters. (a) Gaussian chirped electric fields ($\omega_{0}=0.6m$, $b=0.6$, and $\omega_{m}=1000/m$), (b) frequency-modulated chirped electric fields ($\omega_{0}=0.6m$, $b_{1}=10$, and $\omega_{n}=0.06m$), (c) linearly chirped electric fields ($\omega_{0}=0.55m$ and $b_{2}=5\times10^{-4}m^2$), (d) quadratically chirped electric fields ($\omega_{0}=0.6m$, $\beta_{1}=0.1$, and $\beta_{2}=0.01$), and (e) sinusoidally chirped electric fields ($\omega_{0}=0.6m$, $b_{5}=1.0$, and $b_{6}=1.6$). The peak positions are also marked in the figures. Other electric field parameters are $E_{0}=0.1E_{\mathrm{cr}}$ and $\tau=20/m$. \label{fig:dt}}
\end{figure}

In Fig. \ref{fig:ntmax}, the number density of created EP pairs as a function of chirped parameters and the carrier angular frequency for different chirped electric fields is plotted.
It should be noted that the fixed parameters, $\omega_0=0.6m$ in (a), $\beta_1=0.1$ in (c), and $b_6=1.6$ in (d), are selected to maximize the final number density.
From Fig. \ref{fig:ntmax}(a), one can see that the number density increases with the increase of chirp parameters $b_{1}$ and $\omega_{n}$.
Furthermore, the impact of $b_{1}$ is more important than that of $\omega_{n}$.
Specifically, in regions where the chirp parameter $b_{1}$ takes on large values, the change of $\omega_{n}$ has a significant influence on the number density.
However, in regions with small values of the chirp parameter $b_{1}$, the change of $\omega_{n}$ has a weak effect on the number density.
For this frequency-modulated chirped electric field, the maximum value of the number density is $n^{\mathrm{max}}_\mathrm{F}=1.064\times10^{-5}m^{-3}$.
The corresponding electric field parameters are $\omega_{0}=0.6m$, $b_{1}=10$, and $\omega_{n}=0.06m$.
Figures. \ref{fig:ntmax}(b) and (c) show the situations in linearly chirped and quadratically chirped electric fields, respectively.
The number density in both cases has a similar trend with changes in chirp parameters and the carrier angular frequency.
As the chirp parameters grow, the number density can increase by multiple orders of magnitude.
With the increase of $\omega_{0}$, the number density generally increases as well.
However, when $\omega_{0}$ is in the range of $0.3m$ to $0.6m$, the increase in the carrier angular frequency does not always have a positive impact on the number density.
This can be well explained using Fig. \ref{fig:b0t}.
From Fig. \ref{fig:b0t}, we see that the number density does not monotonically increase with the carrier angular frequency.
It oscillates near the thresholds of multiphoton pair production, see the peaks at $\omega_0=0.36m, 0.43m$, and $0.54m$.
Moreover, we find that the variation of the number density with the carrier angular frequency for a given chirp parameter in Figs. \ref{fig:ntmax}(b) and (c) is almost the same as Fig. \ref{fig:b0t}.
This is because although the carrier angular frequency is affected by the chirp parameters of these two chirped fields, this effect is too weak and can be almost ignored.
This can be seen from Figs. \ref{fig:dt}(c) and (d).
In these figures, there is still only one peak on each frequency spectrum and the carrier angular frequency has hardly changed.
Therefore, the maximum value of the number density for linearly chirped electric fields is almost the same as that for quadratically chirped electric fields. For example, the former is $n^{\mathrm{max}}_\mathrm{L}=2.766\times10^{-8}m^{-3}$ and the latter is $n^{\mathrm{max}}_\mathrm{Q}=2.772\times10^{-8}m^{-3}$. The associated electric field parameters are $\omega_{0}=0.55m$, $b_{2}=5\times10^{-4}m^2$ and $\omega_{0}=0.6m$, $\beta_{1}=0.1$, $\beta_{2}=0.01$, respectively.
Fig. \ref{fig:ntmax}(d) shows the variation of the number density with chirp parameters and the carrier angular frequency for sinusoidally chirped electric fields. We find that when the frequency $\omega_{0}$ is taken with a small value, the change in the chirp parameter $b_{5}$ has a weak effect on the particle pair production. However, when the frequency $\omega_{0}$ is taken with a large value, the number density undergoes significant changes with the chirp parameter $b_{5}$. In addition, both $b_{5}$ and $\omega_{0}$ can enhance the production of particle pairs. The maximum value of the number density in this chirped electric fields is $n^{\mathrm{max}}_\mathrm{S}=1.698\times10^{-2}m^{-3}$. The corresponding field parameters are $\omega_{0}=0.6m$, $b_{5}=1$, and $b_{6}=1.6$.

The preceding discussion indicates that the number density of created EP pairs can be significantly improved by introducing five different chirped electric fields with appropriate field parameters.
Comparing the maximum values of the number density for five chirped electric fields in ascending order, we obtain $n^{\mathrm{max}}_\mathrm{L}<n^{\mathrm{max}}_\mathrm{Q}
<n^{\mathrm{max}}_\mathrm{F}<n^{\mathrm{max}}_\mathrm{G}
<n^{\mathrm{max}}_\mathrm{S}$.
Consequently, the maximum value of the number density of created EP pairs for five chirped electric fields with the field parameters $E_{0}=0.1E_{\mathrm{cr}}$, $\omega_{0}=0.1\sim0.6m$, and $\tau=20/m$ is $1.698\times10^{-2}m^{-3}$.
The associated chirped electric field is the sinusoidally chirped electric field (\ref{eq:filed5}).
The optimal field parameters is $\omega_{0}=0.6m$, $b_{5}=1$, and $b_{6}=1.6$.

To further explain the above results, we present the frequency spectra of five different chirped electric fields at their respective optimal field parameters, as shown in Fig. \ref{fig:dt}.
Firstly, we find that for each chirped electric fields the maximum value of the number density is almost always obtained at the maximum carrier angular frequency ($\omega_0=0.6m$) and the critical chirp parameters.
This can be understood from Fig. \ref{fig:b0t}. In Fig. \ref{fig:b0t}, it shows that the number density of created EP pairs generally increases with the increase of the carrier angular frequency although there are some local oscillations.
So, the larger the carrier angular frequency, the more favorable it is for the production of particle pairs.
There are similar results for chirp parameters. The critical chirp parameters is used to modulate the electric field to obtain the maximum field frequency.
Of course, there are also a few exceptions, such as for linearly chirped electric fields, the carrier angular frequency ($\omega_0=0.55m$) at which the number density reaches its maximum value is not its maximum value.
This is because the number density at $\omega_0=0.55m$ is larger than that at $0.6m$, see Fig. \ref{fig:b0t}.
The above discussion is about the optimal field parameters for each chirped electric field.
Below, we will explain why the number density of EP pairs created by a sinusoidally chirped electric field is the largest among five different chirped electric fields.
As can be seen from Figs. \ref{fig:dt}(a)-(d), the main frequencies for Gaussian chirped, frequency-modulated chirped, linearly chirped, and quadratically chirped electric fields are $1.200m$, $1.103m$, $0.550m$, and $0.599m$, respectively.
In these chirped fields, EP pairs are mainly produced by absorbing several photons with the aforementioned frequencies.
Based on the relation between the number density and the field frequency shown in Fig. \ref{fig:b0t}, we can simply infer that the number density for these four chirped fields satisfies $n^{\mathrm{max}}_\mathrm{Q}<n^{\mathrm{max}}_\mathrm{L}
<n^{\mathrm{max}}_\mathrm{F}<n^{\mathrm{max}}_\mathrm{G}$.
However, since the main frequencies of the quadratically chirped and linearly chirped fields are relatively close and the width of the main frequency peak for the former field is larger than that for the later, the number density for a linearly chirped field may not necessarily be greater than that for a quadratically chirped field.
This analysis is also applicable to the frequency-modulated chirped and Gaussian chirped fields.
Thus, it is not surprising that we obtained the result of $n^{\mathrm{max}}_\mathrm{L}<n^{\mathrm{max}}_\mathrm{Q}
<n^{\mathrm{max}}_\mathrm{F}<n^{\mathrm{max}}_\mathrm{G}$.
Finally, we consider the number density for the sinusoidally chirped electric field. From \ref{fig:dt}(e), one can see that there are many peaks in the spectrum and many peaks have frequencies greater than $m$ or even $2m$.
Hence, according to Fig.\ref{fig:b0t}, where the number density experiences a jump by orders of magnitude after the frequency exceeds $2m$,  the sinusoidally chirped electric field can significantly improve the production of EP pairs and obtain the highest number density.

\subsubsection{Maximizing the enhancement factor}
In the previous subsection, we found the maximum number density and corresponding optimal chirped field in pair production. However, sometimes we are more interested in how much the number density increases for a chirped field compared to that for a chirp-free field. So, in this section, we will study the enhancement factor (the ratio of the number density for the electric field with and without chirp, i.e., $n/n_0$) and find its maximum value. The chirped electric field and its optimal field parameters corresponding to the maximum enhancement factor are also explored in the field parameters $E_{0}=0.1E_{\mathrm{cr}}$, $\omega_{0}=0.1\sim0.6m$, and $\tau=20/m$.

\begin{figure}[!ht]
\centering
\subfigure{\includegraphics[width=0.48\textwidth]{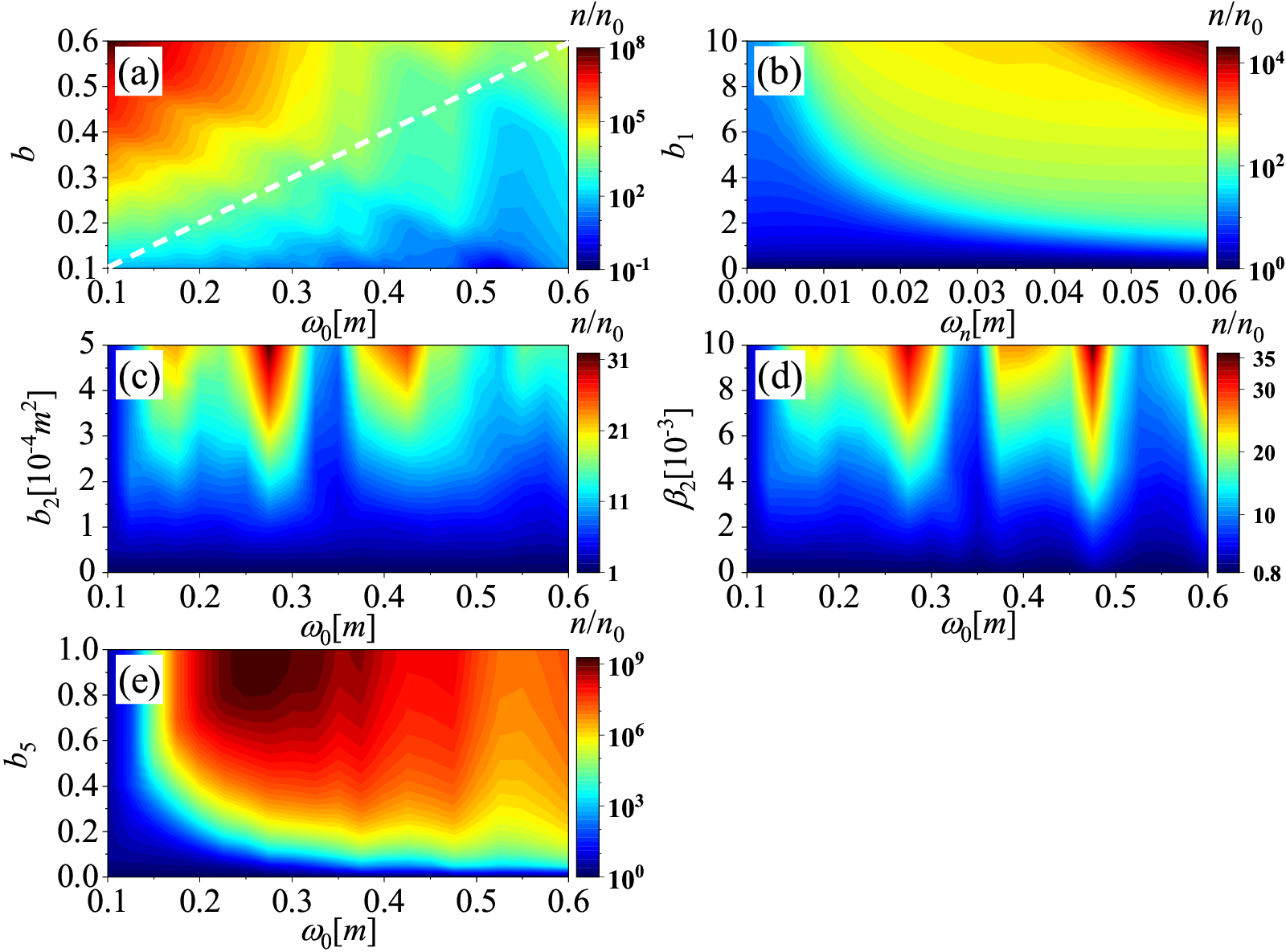}}
\caption{The enhancement factor as a function of chirp parameters and the carrier angular frequency for different chirped electric fields. (a) Gaussian chirped electric fields with $\omega_{m}=1000/m$, (b) frequency-modulated chirped electric fields with $\omega_{0}=0.6m$, (c) linearly chirped electric fields, (d) quadratically chirped electric fields with $\beta_{1}=0.1$, and (e) sinusoidally chirped electric fields with $b_{6}=1.6$. Other electric field parameters are $E_{0}=0.1E_{\mathrm{cr}}$ and $\tau=20/m$. \label{fig:cp}}
\end{figure}

Fig. \ref{fig:cp}(a) presents the enhancement factor as a function of the chirp parameter $b$ and the carrier angular frequency $\omega_0$ for a Gaussian chirped electric field.
Similar to Sec. \ref{sub:1}, we only focus on the date on the right half side of the white dashed line.
In the figure, the enhancement factor monotonically increase with $b$ while changes with $\omega_{0}$ in a fluctuating state.
Thus, within the range of field parameters we used, choosing a larger chirp parameter $b$ and a suitable carrier angular frequency $\omega_{0}$ may have a greater enhancement in the production of EP pairs.
The simulation calculation results also confirm this. By choosing $b=0.6$, $\omega_{0}=0.6m$, and $\omega_{m}=1000m$, the enhancement factor reaches its maximum value, i.e., $n/n_0=1.50\times10^{4}$.
In Fig. \ref{fig:cp}(b), it shows that the increase of the chirp parameters $b_{1}$ and $\omega_{n}$ can significantly enhance the production of EP pairs.
Specifically, within the region where the chirp parameter $b_{1}$ is small, the change of the chirp parameter $\omega_{n}$ have a less noticeable influence on pair production.
However, in the region where the chirp parameter $b_{1}$ is larger, the increase of $\omega_{n}$ exhibits a remarkable impact on pair creation.
That is, the large chirp parameters $b_{1}$ and $\omega_{n}$ are more favorable for enhancing the production of EP pairs.
For the frequency-modulated chirped electric field, the maximum enhancement factor field is $n/n_0=1.34\times10^{4}$. The corresponding electric field parameters are $b_{1}=10$, $\omega_{n}=0.06m$, and $\omega_{0}=0.6m$.
One can see that this enhancement factor is comparable to that for the Gaussian chirped electric field.
The enhancement factors as a function of chirp parameters and the carrier angular frequency for linearly chirped and quadratically chirped electric fields are shown in Figs. \ref{fig:cp}(c) and (d), respectively.
The two figures exhibit similar pattern.
In both cases, an increase in the chirp parameters is beneficial for the production of EP pairs.
With the increase of the chirp parameters, the effect of the carrier angular frequency on pair production becomes more pronounced.
For large chirp parameters, the enhancement factors show periodic oscillations with the variation of carrier angular frequency: first becoming stronger, then weaker, then stronger, and then weaker again.
These two chirped electric fields have complex effects on pair production.
In fact, their enhancement effects are not very good.
For the linearly chirped electric fields, the enhancement factor reaches its maximum value of $31.63$ at $\omega_{0}=0.275m$ and $b_{2}=5\times10^{-4}m^2$.
For the quadratically chirped electric fields, the enhancement factor achieves its maximum value of $35.79$ at  $\omega_{0}=0.475m$, $\beta_{1}=0.1$, and $\beta_{2}=0.01$.
Both of them are not very effective in promoting the production of EP pairs.
Fig. \ref{fig:cp}(e) depicts the variation of enhancement factor with chirp parameters and the carrier angular frequency for sinusoidally chirped electric fields.
The chirp parameter $b_6$ is set to significantly enhance the production of EP pairs.
Similar to Figs. \ref{fig:cp}(a), (c), and (d), the enhancement factor in Fig. \ref{fig:cp}(e) monotonically increases with the chirp parameter $b_5$ for a given carrier angular frequency $\omega_0$, while it oscillates with the carrier angular frequency for a given chirp parameter.
Therefore, we can conclude that a larger chirp parameter and an appropriate carrier angular frequency can greatly enhance pair production.
The simulation results in Fig. \ref{fig:cp}(e) prove this point. The enhancement factor for the sinusoidally chirped electric field reaches its maximum value of $1.74\times10^{9}$ by selecting the carrier angular frequency $\omega_{0}=0.275m$ and the chirp parameters $b_{5}=1$ and $b_{6}=1.6$.
Finally, we find that the maximum enhancement factors for these five chirped electric fields satisfy the same magnitude relationship as the maximum number density discussed in the previous section.
The sinusoidally chirped electric field yields the largest enhancement factor among all chirped electric field configurations.

Unlike the case of maximizing particle yield, the above research results indicate that the maximum enhancement factor is not always achieved at the maximum carrier angular frequency and chirp parameters.
For instance, the carrier angular frequencies corresponding to the maximum enhancement factor for the linearly polarized, quadratically chirped, and sinusoidally chirped electric fields are $0.275m$, $0.475m$, and $0.275m$, respectively.
This finding provides us with the possibility to more efficiently enhance the production of EP pairs by optimizing the chirped electric field parameters.

\begin{figure}[!ht]
\centering
\includegraphics[width=0.48\textwidth]{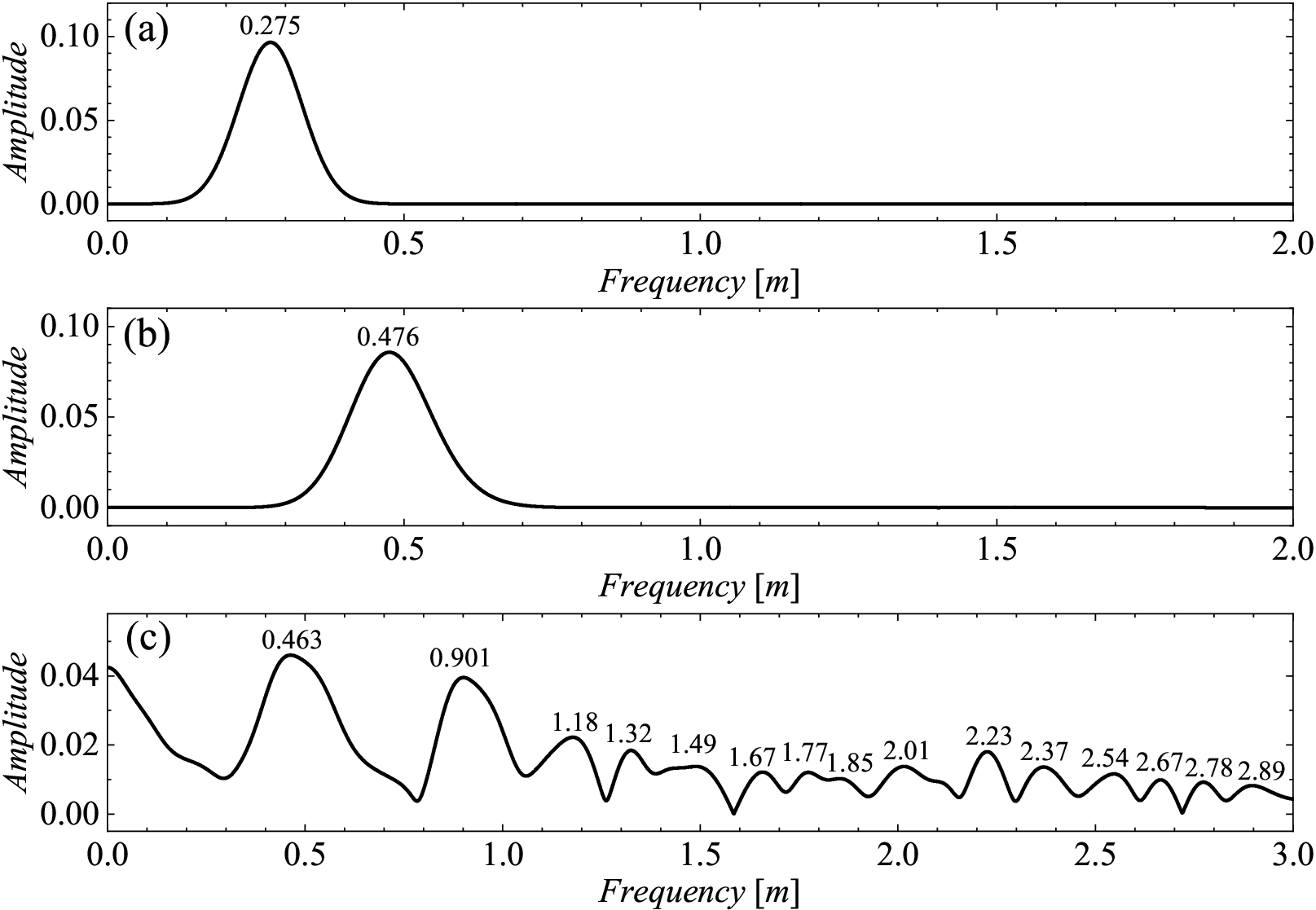}
\caption{The frequency spectra of different chirped electric fields at the optimal field parameters.  (a) Linearly chirped electric field ($\omega_{0}=0.275m$ and $b_{2}=5\times10^{-4}m^2$), (b) quadratically chirped electric field ($\omega_{0}=0.475m$, $\beta_{1}=0.1$, and $\beta_{2}=0.01$), (c) sinusoidally chirped electric field ($\omega_{0}=0.275m$, $b_{5}=1.0$, and $b_{6}=1.6$). The peak positions are also displayed in the figures. Other electric field parameters are $E_{0}=0.1E_{\mathrm{cr}}$ and $\tau=20/m$. \label{fig:max}}
\end{figure}

By further analyzing the oscillation of the number density with the carrier angular frequency in Figs. \ref{fig:cp}(a), (c), (d), and (e), we find that the peak and valley positions of the oscillations are very similar.
Specifically, the peak positions are generally around $\omega_0=0.275m$, $0.475m$, and $0.6m$, while the valley positions are near $\omega_0=0.35m$, $0.45m$, and $0.55m$. These carrier angular frequencies are exactly close to the peak and valley positions in Fig. \ref{fig:b0t}.
It is worth noting that for the case of Fig. \ref{fig:cp}(b), if we fix the chirp parameters and consider the variation of the enhancement factor with the carrier angular frequency, a similar phenomenon will also occur.
This indicates that the above phenomenon is universal.
The reason for this phenomenon is that the introduction of chirp changes the relation between the number density and the carrier angular frequency, leading to a shift in the peak and valley positions in Fig. \ref{fig:b0t}.
Thus, the difference in the number density between chirped and chirp-free cases becomes smaller near the peak regions, corresponding to a smaller enhancement factor.
Conversely, in the valley regions, the number density exhibits greater disparity, resulting in a larger enhancement factor.
Since the effect of the chirp on the relation between the number density and the carrier angular frequency is very complex,
it is difficult to quantitatively explain why the maximum enhancement factor appears at the optimal field parameters.
However, we can qualitatively understand the result.
To do so, we plot the frequency spectra of different chirped electric fields at their respective optimal parameters in Fig. \ref{fig:max}.
Note that the frequency spectra of Gaussian chirped and frequency-modulated chirped electric fields are shown in Figs. \ref{fig:dt}(a) and (b), respectively.
For Gaussian chirped electric fields with $\omega_m=1000/m$, the main frequency becomes $\omega_0+b$.
If we only consider the effect of chirp parameters on the carrier angular frequency while neglecting its influence on the spectral width and local field strength, then according to Fig. \ref{fig:b0t}, we can calculate the ratio of the number density of EP pairs created by absorbing photons with main frequency versus carrier angular frequency.
The result shows that the ratio reaches its maximum at $\omega_0=0.6m$ and $b=0.6m$, which is the same as the result in Fig. \ref{fig:cp}(a).
For frequency-modulated chirped electric fields, due to the constraint $b_1\omega_n\leq\omega_0$, the larger the carrier angular frequency for a given $b_1$, the larger the maximum value of the chirp parameter $\omega_n$.
Then the main frequency shifts towards higher frequencies, and more peaks will appear at the same time.
Pair production will be significantly enhanced.
This does not mean that the higher the carrier angular frequency, the greater the enhancement factor.
The enhancement factor oscillates as the carrier frequency increases.
However, when $\omega_0=0.6m$, the number density itself reaches a minimum at this frequency, see Fig. \ref{fig:b0t}.
Meanwhile, the main frequency of the chirped field shifts to $1.1m$, which greatly improves the number density.
As a result, the ratio of the number density between chirped and chirp-free cases attains its maximum value.
For linearly chirped and quadratically chirped electric fields, the optimal field parameters corresponding to the maximum enhancement factors can not be understood by the increase of carrier angular frequency, because the main frequency is almost identical to the carrier angular frequency, see Figs. \ref{fig:max}(a) and (b).
The number density of EP pairs created by these two chirped fields
is mainly affected by the increase of spectral width and local field strength.
Thus, the enhancement effect is quite limited and has a certain impact on the peaks near multiphoton absorption thresholds.
Overall, the enhancement factor is relatively small.
It may be smaller near the peaks and slightly larger near the valleys.
For sinusoidally chirped electric fields, there are many peaks in the frequency spectrum, see Fig. \ref{fig:dt}(e) and Fig. \ref{fig:max}(c).
Even for smaller carrier angular frequencies, the frequency spectrum contains higher frequency components, even those exceeding the single photon absorption threshold $2m$, so pair production is greatly enhanced by absorbing these high-frequency photons.
Furthermore, since the number density of EP pairs produced by the electric field with smaller carrier angular frequencies is very low, the enhancement factor at small carrier angular frequencies is relatively large.
Of course, for particularly small carrier angular frequencies, the enhancement factor is small due to the absence of high-frequency components in their chirped fields.
For larger carrier angular frequencies, although there are more high-frequency components in the chirped fields, the increase in number density is not so significant after the frequency reaches a certain value.
For instance, when the frequency is greater than $2m$, the number density changes very little with increasing frequency, see Fig. \ref{fig:b0t}.
Additionally, the number density at a large carrier angular frequency is more larger than that at a small one.
Therefore, the enhancement factor at large carrier angular frequencies is also smaller.
Ultimately, the maximum enhancement factor is expected to occur at an intermediate carrier angular frequency-neither too small nor too large.

\section{Conclusion and outlook }
\label{sec:four}

In summary, we first investigated the momentum spectrum and number density of EP pairs created by a Gaussian chirped electric field.
The momentum spectrum exhibits a distinct interference pattern, which is related to EP pair production by the absorption of photons with different frequencies.
The number density increases monotonically with the chirp parameters while oscillates with the carrier angular frequency.
Within a reasonable range of field parameters, the number density can be improved by $4$ orders of magnitude compared to that for a chirp-free field.
This provides us with a new way to enhance EP pair production by chirped electric fields.
We also examined the number density of EP pairs produced by the other four types of chirped fields: frequency-modulated chirped, linearly chirped, quadratically chirped, and sinusoidally chirped  electric fields.
By analyzing the variation of the number density with chirp parameters and the carrier angular frequency, it is found that the number density for each chirped fields achieves its maximum at the maximum chirp parameters and maximum carrier angular frequency.
Their maximum number density satisfies the following relationship: sinusoidally chirped field $>$ Gaussian chirped field $>$ frequency-modulated chirped field $>$ quadratically chirped field $>$ linearly chirped field.
The maximum number density of EP pairs created by sinusoidally chirped electric fields is the highest.
By studying how the enhancement factor varies with chirp parameters and carrier angular frequency, we found that the maximum enhancement factors for five chirped electric fields satisfy the same ranking relationship as the maximum number density.
The maximum enhancement factor for sinusoidally chirped electric fields is the largest.
The number density for this chirped field can be improved $9$ orders of magnitude compared to that for the chirp-free field.
These findings further deepen our understanding of EP pair production by chirped fields, identify the optimal chirped field that can significantly enhance pair production, and will also provide important theoretical references for experimental verification of EP pair production.

In our current investigation, we have considered five types of chirped electric fields with a certain constraint on the range of field parameters.
Therefore, further research is needed to determine the results for more types of chirped fields with broader field parameters.
In addition, we only considered the case of spatially uniform and time-varying chirped electric fields.
How spatial inhomogeneities might affect the outcomes presents another valuable direction for future research.

\begin{acknowledgments}
The work is supported by the National Natural Science Foundation of China (NSFC) under Grants No. 11974419 and No. 11705278, the Strategic Priority Research Program of Chinese Academy of Sciences (Grant No. XDA25051000, XDA25010100), and the Fundamental Research Funds for the Central Universities (No. 2023ZKPYL02, 2025 Basic Sciences Initiative in Mathematics and Physics).
\end{acknowledgments}

\end{document}